# Cell Lists Method Based on Doubly Linked Lists for Monte Carlo Simulation

Shaoyun Wang, Chaohui Tong*

tongchaohui@nbu.edu.cn

*School of Physical Science and Technology, Ningbo University, Ningbo, Zhejiang, 315211, China*

**Abstract** A cell lists method based on doubly linked lists and with complexity $O(N)$ is developed for particle deletion and insertion in reaction ensemble Monte Carlo simulation. Because the random move in Metropolis algorithm can be reduced to particle deletion at old position and particle insertion at new position, so this method can be also used in Metropolis algorithm. In addition, nonlocal move, common in Monte Carlo simulation of polymers, such as kink-jump, pivot, reptation move and the retrace and regrow of chains in configurational biased Monte Carlo often cause the failure of Verlet lists method because the large displacement in these nonlocal moves will exceed Verlet cutoff radius. So we also use cell lists method based on doubly linked lists to achieve nonlocal move in this study.

## 1. Introduction

Monte Carlo (MC) and Molecular Dynamics (MD) are two main methods that involve solving many-body problem based on statistical mechanics and classic mechanics to obtain physical properties of macro matters. In MC and MD simulation, the most time-consuming part is the calculation of long-range and short-range interactions. For calculation of short-range interaction energy of the system with particles $N$, the most direct method is to sum $N(N-1)/2$ pairs interactions energy by double loops directly. The computational complexity of this method scales as $N^2$, which is impractical for large systems. To speed up the computation, two methods with computational complexity $\mathcal{O}(N)$ are developed in MD [1-3]. The first method is cell lists (or linked lists) method. In this method, the system is divided into many cells first, then particles in each cell are linked by the linked-list technology [1]. When calculating interactions of other particles with particle $i$, only particles in the cell where particle $i$ locate in and neighbor cells of that cell are considered. These neighbor particles is proportional to $\mathcal{O}(1)$, so the complexity of all particles is $\mathcal{O}(N)$. The second one is Verlet lists (or neighbor lists) method. In this method, neighbor particles within a second cutoff radius $r_v$ which is little greater than interaction cutoff radius $r_c$ is prestored in a list which is noted as Verlet list. When calculating the interaction, only the particles within radius $r_v$ is considered and they scale as $\mathcal{O}(1)$, so the complexity of the whole system is $\mathcal{O}(N)$. Meanwhile, Verlet lists is updated by using cell lists method whose complexity of $\mathcal{O}(N)$ [1]. Similarly, there should be two corresponding methods in Monte Carlo simulation. The Verlet lists method is existed [1], however, how to update cell lists for each random move with complexity $\mathcal{O}(N)$ in Monte Carlo has not been reported.

Generally, Verlet lists method is more efficient than the pure cell lists method [1], so it seems there is no necessity to develop the cell lists method in Monte Carlo simulation. However, there are some special move which we call nonlocal move in Monte Carlo simulation and Verlet lists fails for these move. The first example is Monte

Carlo simulation of chemical reaction by using reaction ensemble Monte Carlo method [2, 4] and of ionization equilibrium of weak acid by using constant pH method [5-6]. In these situation, particle insertion and deletion are involved to simulate reaction and ionization. In prestored Verlet lists, when a new particle $i$ is inserted, the neighbor lists of particles within radius $r_v$ of particle $i$ is needed to add particles $i$ and neighbor list of particle $i$ is also needed to be constructed. But the order number of these neighbor particles of $i$ is difficult to find so it is difficult update the Verlet lists with $\mathcal{O}(1)$. The most easy way is updating all Verlet lists in each particle insertion. However, we need to update $N$ times Verlet lists in each Monte Carlo move which includes $N$ times particle insertion, so the complexity is $\mathcal{O}(N^2)$ which is inacceptable. Similarly, Verlet lists method fails for particle deletion. The second example is Monte Carlo simulation for polymers. As we know, the equilibrium time Metropolis algorithm for polymer simulation is extremely long [1,7]. To speed up the simulation, we can use nonrealistic moves such as reptation, kick-jump, pivot move [8-9] or use configurational-bias Monte Carlo (CBMC) method [1,10]. For reputation move, one terminal particle is deleted and is added on another terminal particle. Just like the case in chemical reaction, particle insertion and deletion are involved which cause the failure of Verlet lists method. For kink-jump and pivot move, the displacement is quite large compared with Monte Carlo move so that some particles will exceed the buffer distance $(r_v - r_c)/2$ and the Verlet lists are needed to be reconstructed. For CBMC move, part of chains are retraced one by one at first and then they are regrown one by one in other positions. Similar to the pivot move, the displacement of particles far away starting particle is large so that Verlet lists need to be updated each time.

One method to solve the problem is the linked neighbor list method developed by Mazzeo [11]. By combining Verlet lists and linked lists, he linked all neighbor pairs in Verlet lists so that the neighbor particles of the deleted or added particle can be found by linked lists. However, the data structure of this method is a two dimensional array where the rows are particles $N$ in the system and the columns are maximum neighbor particles $LNS_{MAX}$. $LNS_{MAX}$ is difficult to evaluate and it is set to a large number for safety. Therefore, the memory of this method is extremely large for dense system. In this study, cell lists method based on doubly linked lists [12] is developed to achieve the update of cell lists with complexity $\mathcal{O}(N)$ for particle insertion and deletion in Monte Carlo simulation. All these moves can be reduced to particle insertion and particle deletion in fact. Therefore, this method is not only suit for random move in Metropolis algorithm but also suit for all the former nonlocal move. In section 2, we give the detailed description the algorithms of this method. In section 3, we check the results by comparing this method with Verlet lists for simulating of Lennard-Jones liquid and compare the efficiency of this method with Verlet lists method.

**2. Methods and algorithms**

2.1 Overview

In this section, we describe the cell lists method detailly and give the corresponding algorithms. At first, we describe the construction of linked lists and the calling for linked lists to calculate the energy. Then, we describe the update of cell lists for particle deletion and insertion. Finally, we use the cell lists method to achieve the Metropolis move and nonlocal move.

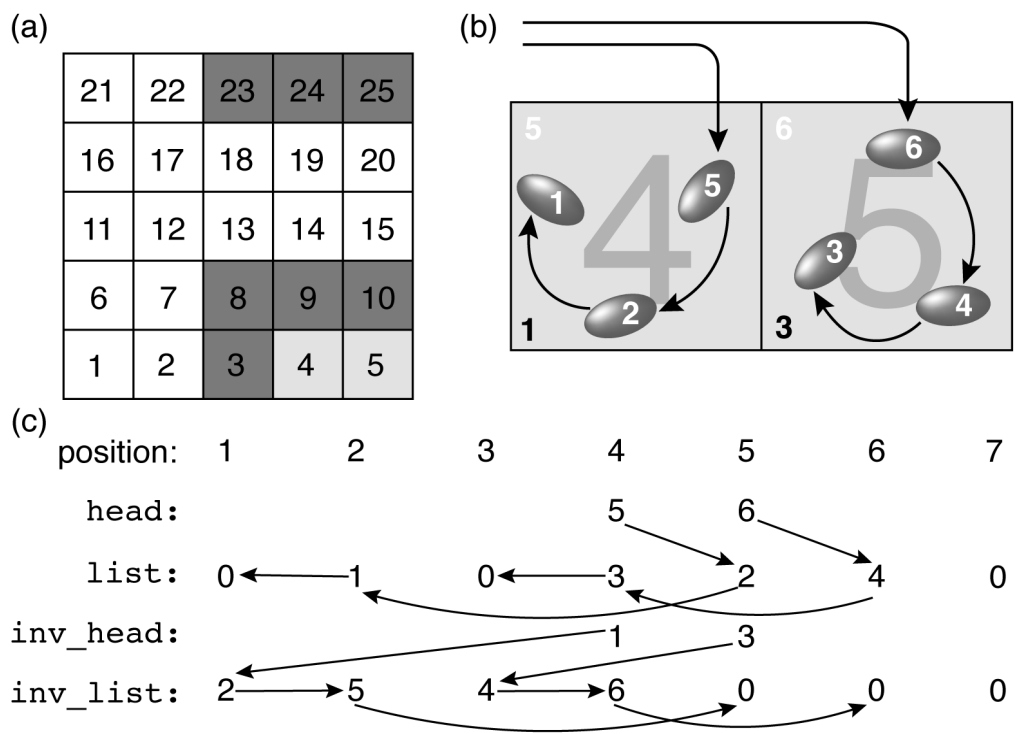

Figure 1. The cell lists method in two dimension. (a) The central box is divided into $s_c \times s_c$ square cells ($s_c = 5$) with side length $r_c$. A molecular in cell 4 may be interacted with molecules only in cell 4 and other 8 neighbor cells which is noted by shaded color. Periodic condition is used in this condition. (b) We construct the cell lists of cell 4 and 5 which includes particle 1, 2, 5 and particle 3, 4, 6 respectively. These particles can be connected by the arrows. The white numbers at northwest of two cells are the head numbers of forward linked lists, and black numbers at southwest of two cells are the head numbers of backward linked lists. (c) The `head` array whose length is cells of the system is the head of chains array. For example, head particle in cell 4 is particle 5, so `head(4)=5`. The `list` array whose length is particles in the system is forward linked list whose element points to next index.

For example, particle 5 points to particle 2 and points to particle1 so `list(5)=2`, `list(2)=1`. Particle 1 is the end particle and it points to 0, so `list(1)=0`. Similarly, `inv_head` array stores the head particle of backward linked list and `inv_list` array is backward linked list.

---

**Algorithm 1.** The construction of cell lists

This program is used to construct linked list `list` and head of chains array `head`. For convenience, the codes are programmed in one dimension with periodic condition. `x` is the position array with size of particles `n` in the system. `icel` is an index number to note the serial number of cell. `sc` is cells in x axis and `rc` is cutoff radius and is also side length of each cell.

```
real,    dimension(n)      :: x
integer, dimension(n)      :: list
integer, dimension(0:sc-1) :: head
integer                    :: icel

head = 0
do i = 1, n
  icel = int(x(i)/rc)
  list(i) = head(icel)
  head(icel) = i
end do
```

---

2.2 Construction of cell lists and using cell lists to calculation energy

Now we turn to the construction of linked lists and inverse linked lists and the schematic diagram is shown in Fig. 1. The central box can be divided into $s_c \times s_c$ square cells with side length $r_c$ and each cell is denoted by an index. Particles in each cell can be linked by arrows from the first particle to the end particle. This link of particles can be achieved by a linked array `list` whose index is current particle number and the element point to the next index which is the next particle number. The starting particle number in each cell is stored in a `head` array with a size of the number of cells in the system. The end particle in each cell cannot points to other particle so it points to 0. The algorithm of the construction of linked list is shown in Algorithm 1. However, for particle deletion in linked list, we need to find the former particle of the deletion particle so we need to use the inverse linked list. The linked list (or forward linked list) and the inverse linked list (or backward linked list) construct the doubly linked lists which is the key to achieve particle deletion. The construction of inverse linked list is similar to the construction of linked list except the order of the loop and it is shown in Algorithm 2.

---

**Algorithm 2.** The construction of inverse cell lists

This program is used to construct inverse linked list `inv_list` and head of chains array `inv_head`. Variables undefined is same as them in previous Algorithms.

```
integer, dimension(n)     :: inv_list

integer, dimension(0:sc-1) :: inv_head

head = 0

do i = n, 1, -1
```

```
  icel = int(x(i)/rc)

  inv_list(i) = inv_head(icel)

  inv_head(icel) = i
end do
```

Using cell lists method, we can achieve the calculation of energy with complexity $\mathcal{O}(N)$. At first, we loop all particles, then calculate the energy of each particle with the neighbor particles, and the summation of energy of each particle is total energy. Going through particles in each cell can be achieved by the following process. Take the cell 4 as an example, the starting particles 5 of cell 4 is stored in `head(4)`, or `head(4)=5`. Then, the next number of particle 5 can be found in the 5th element in array `list` and it is 2, or `list(5)=2`. The next number of particle 2 can be found in the 2th element in array `list` and it is 1, or `list(2)=1`. Finally, the 2th element in array `list` is 0 which means it is the terminal particle of this cell. Similarly, we can go through particles in the neighbor cells. The particles in the neighbor cells scales as $\mathcal{O}(1)$ and the loop of all particles scales as $\mathcal{O}(N)$, so the total complexity of calculation energy is $\mathcal{O}(N)$. The codes of calculation energy of the system is shown in Algorithm 4.

**Algorithm 3.** Computation of energy by cell lists

This program is used to calculate energy of the system by cell lists method. Periodic condition is used. Variables undefined can be referred to previous Algorithms.

```
real :: E                              ! Energy

E = 0

do i = 1, n
```

```
   icel = int(x(i)/rc)              ! current cell index
   do k = -1, 1
     jcel = icel + k                 ! neighbor cell index
     if (jcel > sc) jcel = jcel − sc ! Periodic condition
     if (jcel < sc) jcel = jcel + sc ! Periodic condition
     j = head(jcel)                  ! head particle j
     do while( j /= 0 )
       if ( i /= j ) then
         xij = x(i) − x(j)
         xij = xij − nint(xij / L) * L  ! Periodic condition
         E = E + U(abs(xij))         ! Calculate pair interaction
       end if
       j = list(j)                   ! Point to next particle
     end do
   end do
 end do
```

2.3 Particle deletion

Now we turn to how to update the linked list for particle deletion. For example, we delete particle 2 in cell 4 in Fig. 2, then the previous particle 5 and next particle 1 are linked while particle 4 is excluded. So we just set `list(5)=1` and `inv_list(1)=5` to achieve particle deletion. For general circumstance, if we use singly linked list only, we can find the next particle of the deletion particle $i$ by calling `list(i)`, but cannot find previous particle. Therefore, we cannot link the previous particle to the next particle. To find previous particle, we need to use inverse linked list and the previous particle is `inv_list(i)`. Once the previous particle and next particle are found, we can link the previous particle to next particle in linked list and link the next particle to previous particle in inverse linked list to achieve particle deletion. However, if the deletion particle is the first particle in the cell, there is no previous particle so we need to change the `head` array. In addition, if the deletion particle is the last particle in the cell, there is no next particle so we need to change the `inv_head` array. Moreover, if there is only one particle in the cell, it is not only the first particle but also the last particle so we need to change the `head` array `inv_head` array both. These four cases are considered in the Algorithm 4 to achieve particle deletion. The changes of linked list and inverse linked list of each particle deletion are only two elements, so the complexity scales only $\mathcal{O}(1)$ for each particle deletion and $\mathcal{O}(N)$ for Monte Carlo step.

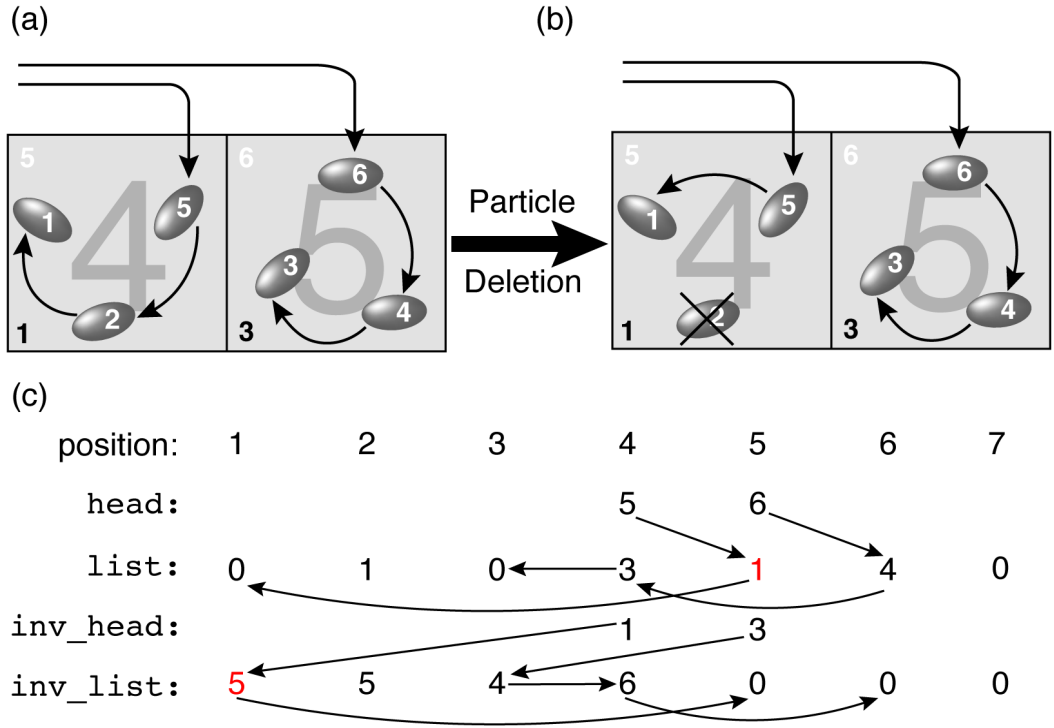

Figure 2. Schematic diagram for particle deletion in cell lists. (a) Original cells and particles, and it is same as the circumstance in Fig. 1. (b) Particle 2 in cell 4 is deleted, so the arrow from previous particle to current particle and arrow from current particle to next particle are broken off and a new arrow from previous particles to next particle is connected. (c) Because particle 5 points to particle 1 now, the 5th index in `list` is become to 1 and the 1st index in `inv_list` is become to 5.

**Algorithm 4.** Change of cell lists after particle deletion

The algorithm describe the change of cell lists `list, inv_list, head, inv_head` after deleting `i`-th particle in `j`-th cell. This algorithm is divided into 4 cases, the first is the deleted particle is in the middle of the linked lists. The second is that it is the starting particle of the linked lists. The third is that it is the end particle of the linked lists, and the final is that it is not only the starting particle but also the end particle of the linked lists.

```fortran
integer                    :: i       ! Particle i is deleted
integer                    :: j       ! in cell j
integer                    :: prei    ! previous particle of i
integer                    :: nti     ! next particle of i
prei = inv_list(i)
nti  = list(i)
if (prei /= 0 .and. nti /= 0) then        ! middle particle
  list(prei) = nti
  inv_list(nti) = prei
elseif ( prei ==0 .and. nti /= 0 ) then  ! starting particle
  head(j) = nti
  inv_list(nti) = prei
elseif ( prei /= 0 .and. nti == 0 ) then ! end particle
  list(prei) = nti
  inv_head(j) = prei
else                                      ! starting and end particle
  head(j) = nti
  inv_head(j) = prei
end if
```

2.4 Particle insertion

Then we consider how to update the linked list for particle insertion. For example, we insert particle 7 in cell 4 in Fig. 3, then particle 7 is linked with the starting particle. So particle 7 becomes the starting particle and a new link between particle 5 and 7 needs to be added. For general circumstance, for example, insert a particle $i$ in cell $j$. There are two cases. When there are particles in the cell, for the inverse linked list, the last particle which is `head(j)` is linked with particle $i$ so `inv_list(head(j))=i`, and particle $i$ is the last particle so `inv_list(i)=0`. For linked list, original starting particle `head(j)` which is now is linked with particle $i$ so `list(i)=head(j)` and the starting particle becomes $i$ now so `head(j)=i`. When there is no particle in the cell which will be inserted, the `inv_head` array is also changed. The both two cases are illustrated in Algorithm 6.

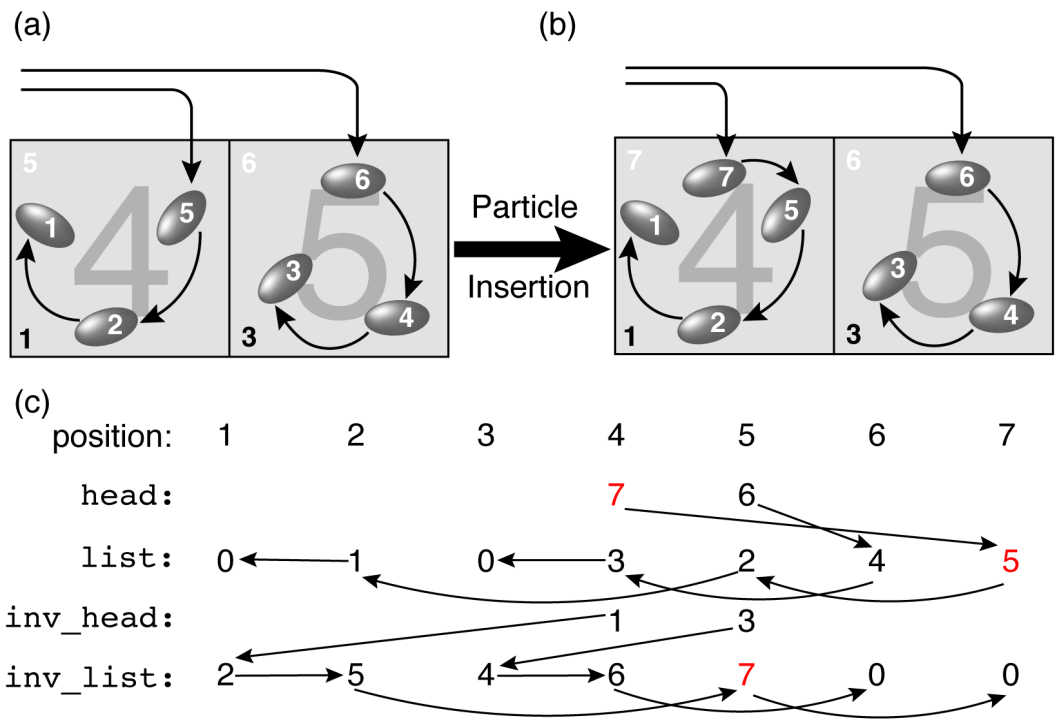

Figure 3. Schematic diagram for particle insertion in cell lists. (a) Original cells and particles, and it is same as the circumstance in Fig. 1. (b) Particle 5 is added in cell 4

and it is connected with the head particle 5. Because the head of cell 4 becomes 7 now, the northwest white number becomes 7. (c) For forward linked list, the 4th index of `head` array becomes 7, and the 7th position of `list` array becomes 5 now. For backward linked list, the 5th index of `inv_list` array becomes 7 and the 7th index of this array becomes 0.

---

**Algorithm 5.** Change of cell lists after particle insertion

The algorithm describe the change of cell lists `list, inv_list, head, inv_head` after inserting `i`-th particle in `j`-th cell. This algorithm is divided into 2 case for the change of inverse linked lists. The first is there is no particle in the cell and the other is there are particles in the cell.

```
Integer                  :: i     ! Particle i is deleted

integer                  :: j     ! in cell j

inv_list(i) = 0                   ! for backward linked lists

if (inv_head(j) /= 0) then        ! particle exists in cell

  inv_list(head(j)) = i

else                              ! no particle in cell

  inv_head(j) = i

end if

list(i) = head(j)                 ! for forward linked lists

head(j) = i
```

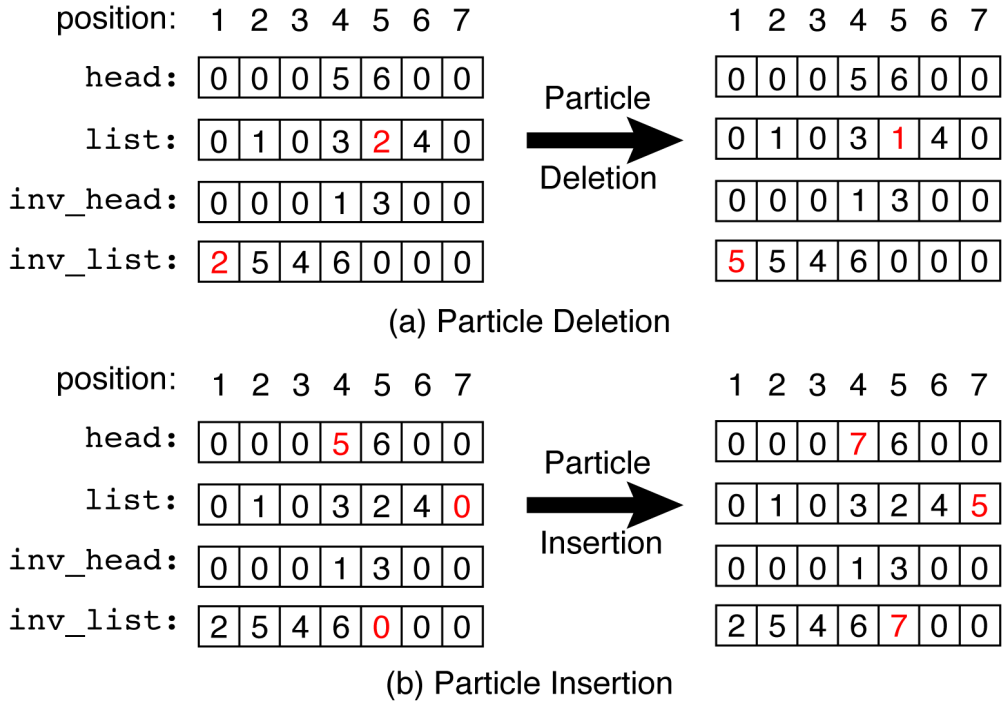

Figure 4. To express the change of arrays in Fig, 2 and Fig. 3 more obviously, we add the box on the arrays. The left arrays in (a) and (b) are the original arrays in Fig. 1. The right arrays in (a) are arrays after particle deletion in Fig. 2. The right arrays in (b) are arrays after particle insertion in Fig. 3.

2.5 Metropolis move and nonlocal move

Once particle insertion and deletion are achieved by cell lists method, we can extent this method to achieve Metropolis move and nonlocal move such as pivot, kink-jump, reputation and CBMC move, which are shown in Fig. 5. For Metropolis move, as is shown in Fig. 6, particle move from cell 4 to cell 5 can be reduced to particle deletion in cell 4 and insertion in cell 5. So we can update linked list and inverse linked list of particle deletion in cell 4 and insertion in cell 5 to achieve the Metropolis move. For pivot move or CBMC move, we can reduce this move to delete particles one by one and insert particles one by one. For reptation move, this can be reduced to particle in one terminal and insert it in the other terminal.

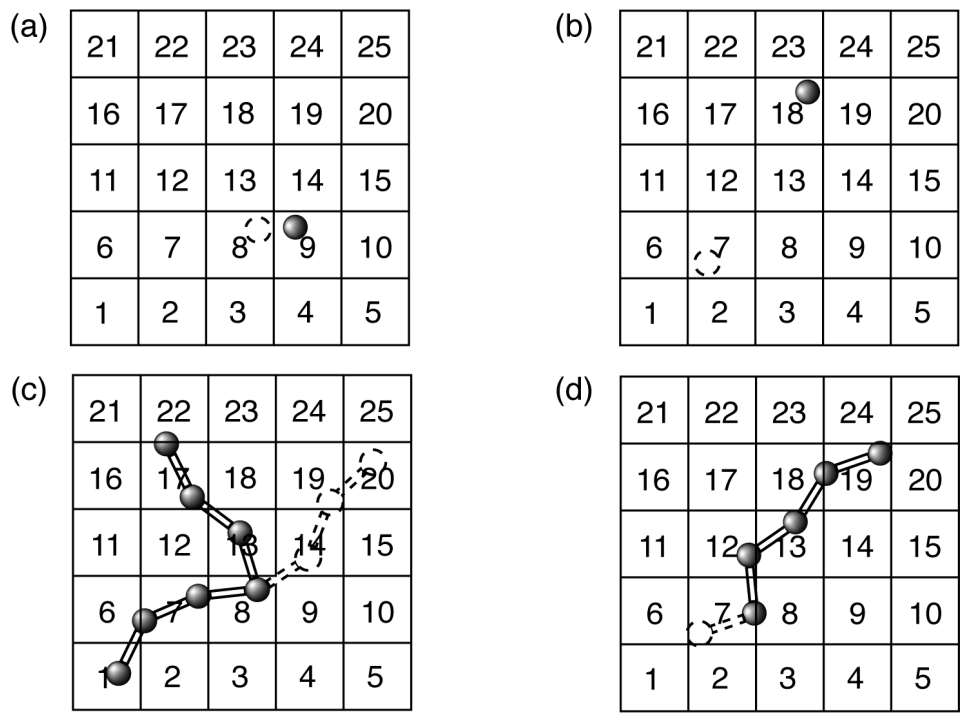

Figure 5. (a) Random move in Metropolis algorithm. (b) Particle deletion and insertion in Monte Carlo simulation of weak acid and base. (c) Pivot move or CBMC move. (4) Reptation move.

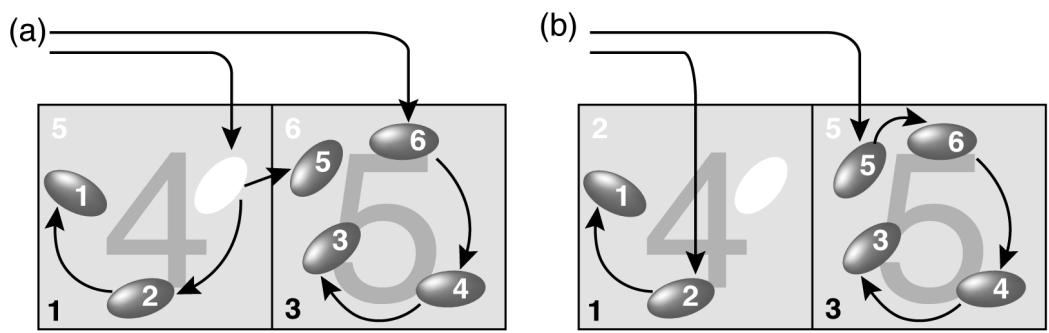

Figure 6. Schematic diagram for particle random move in Metropolis algorithm by cell lists method. (a) Particle 5 move from cell 4 to cell 5. (b) It is equivalent to particle deletion in cell 4 and particle insertion in cell 5.

## 3. Validity and Efficiency

Now we turn to check the validity of cell lists method. We take the simulation of Lennard-Jones fluid in NVT ensemble as the example. Approximate Lennard-Jones potential function is chose as follow [1]:

$$u(r) = \begin{cases} 4\varepsilon\left(\left(\frac{\sigma}{r}\right)^{12} - \left(\frac{\sigma}{r}\right)^{6}\right) & r \leq r_c \\ 0 & r > r_c \end{cases}, \tag{1}$$

where the cutoff radius $r_c$ is set to $2.5\sigma$, $\varepsilon$ and $\sigma$ are energy and length parameters. For convenience, $\varepsilon$ and $\sigma$ are served as energy unit and length unit. Metropolis algorithm is used in simulation and displacement of each move is chose to assure the acceptation ratio to be 30%. Cell lists method and Verlet lists method are used to calculate energy independently. In Verlet lists method, Verlet radius $r_v$ is chose as $3.5\sigma$ in this study. The codes of cell lists method are uploaded to GitHub website https://github.com/wangshaoyun/LJ_Fluid_Cell_list while the codes of Verlet lists are on https://github.com/wangshaoyun/LJ_Fluid_Verlet_list.

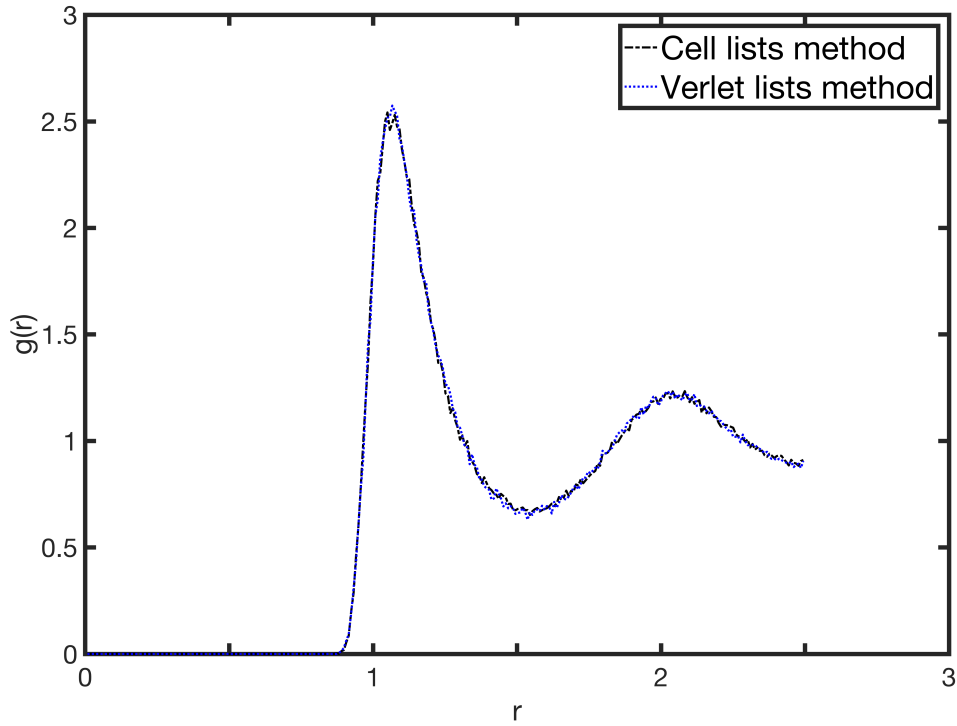

Figure 8. Radial distribution of a Lennard-Jones fluid close to the triple point: $T = 1.5043$ and $\rho = 0.8442$ by cell lists method and Verlet lists method. 500 particles are used in this simulation.

To verify the cell lists method, we compare radial distribution functions simulated by cell lists method and Verlet list method in Fig. 8. As is shown in Fig. 8, both curves are coincident and they are also same as the result in Figure 4.5 in Frenkel's textbook [1]. Therefore, the cell lists method is valid. In addition, we compare the state of equation too. The pressure was calculated using the virial [1]

$$P = \frac{\rho}{\beta} + \frac{vir}{V}, \qquad (2)$$

where $P$ is pressure, $\rho$ is the density, $\beta$ is the reciprocal of temperature, $V$ is the volume of the box and the virial is defined by

$$vir = \frac{1}{3}\sum_i \sum_{j>i} \vec{f}(\vec{r}_{ij}) \cdot \vec{r}_{ij},  \qquad (3)$$

where, $\vec{f}(\vec{r}_{ij})$ is the intermolecular force between particle $i$ and particle $j$. As is shown in Fig. 9, the state of equation simulated by cell lists method and Verlet lists method is coincident too. Moreover, these curves is same as the results in Figure 3.5 in Frenkel's book [1]. Therefore, the cell lists method is right.

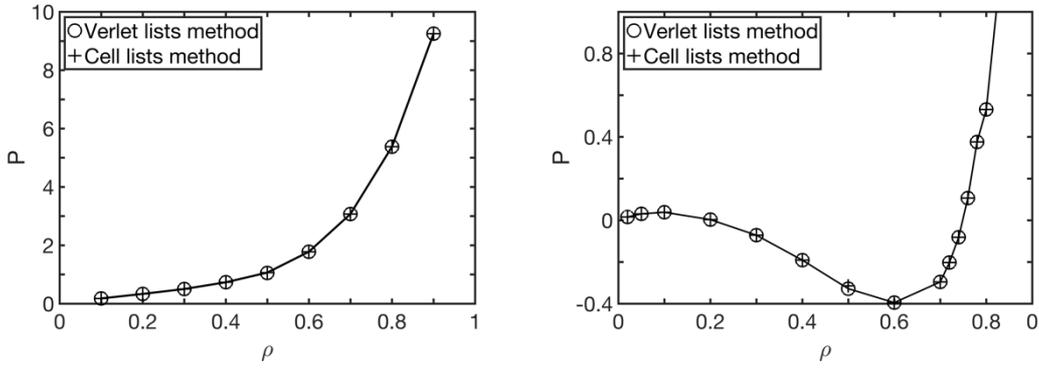

Figure 9. Equation of state of the Lennard-Jones fluid. Temperature of the left figure and right figure are $T = 2.0$ and $T = 0.9$ respectively. 500 particles are used in this simulation.

Finally, we discuss the time complexity and space complexity of cell lists method and Verlet lists method. The simulation is running in Dell Precision T1700 Workstation with Intel ® Xeon (R) CPU E3-1271 v3 @ 3.60GHz. At first, we compare the efficiency of two methods for Lennard-Jones fluid at different density in Fig. 10. The simulation time of cell lists method and Verlet lists method are both proportional to particles at different density, so computational complexity of both methods are $\mathcal{O}(N)$. However, the proportion factor of cell lists is larger than Verlet lists because of more particles are considered in neighbor 27 cells than the particles in the sphere with the Verlet radius $r_v$ [1]. It is the shortcoming of cell lists method compared with Verlet lists method but many technologies are developed to improve the efficiency of cell lists method [11, 13-16]. These technologies all can be extended in this study. For space complexity, both of two method are $\mathcal{O}(N)$, but the size of cell lists is equal to $N$ strictly while the size of Verlet lists is equal to $N \times m_{max}$ where $m_{max}$ is maximum particles in the sphere with cutoff radius $r_v$. $m_{max}$ is difficult to evaluate, so we often set it to be a large number for safety. For dense and large system, the memory that Verlet lists are occupied is very large so the cell lists method is better.

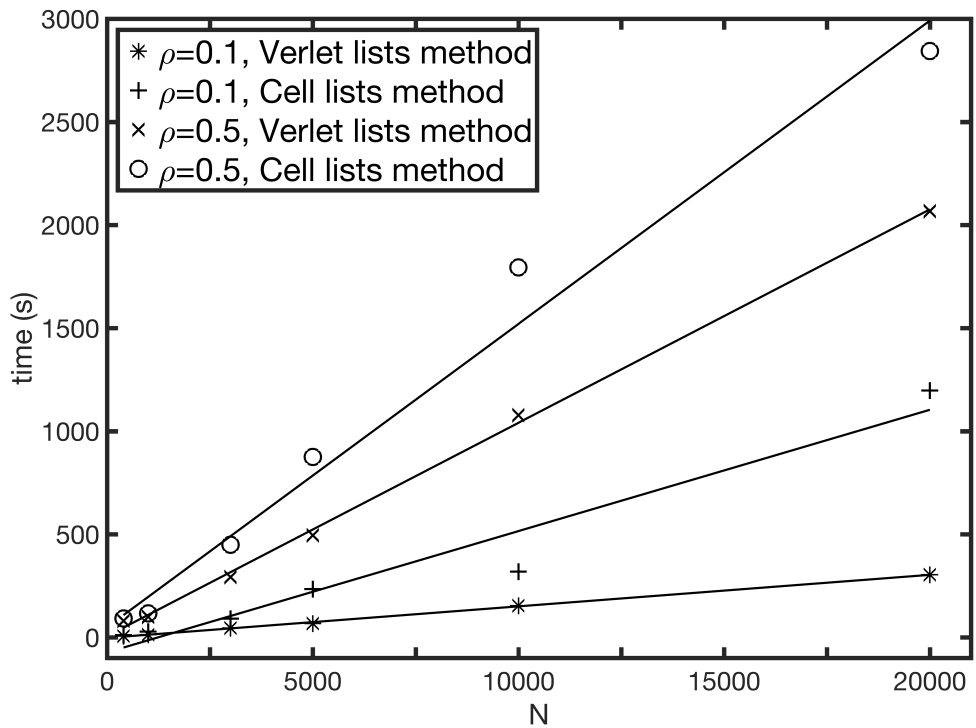

Figure 10. Simulation time for 10000 Monte Carlo steps (MCS) vs. particles of Lennard-Jones fluid at $T = 1.5043, \rho = 0.1$, and $T = 1.5043, \rho = 0.5$ by cell lists method and Verlet lists method.

## 4. Conclusion

In this study, we put forward a cell lists method for particle deletion and insertion in Monte Carlo simulation. In this method, doubly linked lists which not only can find the next particle but also can find previous particle is introduced for particle deletion and insertion. We also described the processes and algorithm of the construction of cell lists, calling for cell lists to calculate energy and update of cell lists after particle deletion and insertion detailly in this paper.

In addition, we extent this method to achieve Metropolis move and nonlocal move such as pivot, kink-jump, reputation and CBMC move since all these moves can be reduced to particle insertion and deletion. Because Verlet lists method fails for these nonlocal moves, so cell lists have good advantage on these situation.

Finally, the time complexity and space complexity of cell lists method are both $\mathcal{O}(N)$, so that it is acceptable for the calculation of short-range interaction. The memory that cell lists are occupied is equal to $N$ strictly, and it is better than Verlet lists. In each particle deletion or insertion, the change of the elements in linked lists is only 2, so it is very fast to update linked list. When using cell lists to calculate energy, particles in neighbor 27 cells are needed to be considered, so they are more than the particles in Verlet cutoff sphere and it costs more time than Verlet. However, many technology are developed earlier, so the efficiency of cell lists method can be improved further.

**Acknowledgements**

The authors thank the financial supports from the National Natural Science Foundation of China (NSFC project 21774067). C. T. acknowledges the support from K. C. Wong